\documentclass{emulateapj}
\usepackage{graphics}
\usepackage{psfig}
\usepackage{ulem}
\usepackage{longtable}

\def\b0540{{B0540-69}}
\def\j0537{{J0537-6910}}
\def\etal{{\it et~al.}}
\def\E{{\sl Einstein Observatory}}

\def\S{{\sl Swift}}
\def\SP{{\sl Swift}.}
\def\X{{\sl RXTE}}

\def\simlt{\lower.5ex\hbox{$\; \buildrel < \over \sim \;$}}

\begin{document}

\title{Discovery of a Spin-Down State Change in the LMC Pulsar B0540-69}

\author{F.\ E.\ Marshall\altaffilmark{1},
L.\ Guillemot\altaffilmark{2,3},
A.\ K.\ Harding\altaffilmark{1},
P.\ Martin\altaffilmark{4},
D.\ A.\ Smith\altaffilmark{5},
}

\email{frank.marshall@nasa.gov}
\altaffiltext{1}{Astrophysics Science Division, NASA Goddard Space Flight Center, Greenbelt, MD 20771, USA}
\altaffiltext{2}{Laboratoire de Physique et Chimie de l'Environnement et de l'Espace
 (LPC2E), CNRS-Universit\'{e} d'Orl\'{e}ans, F-45071 Orl\'{e}ans, France}
\altaffiltext{3}{Station de radioastronomie de Nan\c{c}ay, Observatoire de Paris,
 CNRS / INSU F-18330 Nan\c{c}ay, France}
\altaffiltext{4}{Institut de Recherche en Astrophysique et Plan\'{e}tologie, UPS / CNRS, 
UMR5277, F-31028 Toulouse Cedex 4, France}
\altaffiltext{5}{Centre d'\'{E}tudes Nucl\'{e}aires de Bordeaux Gradignan, 
IN2P3 / CNRS, Universit\'{e} de Bordeaux 1, BP120, F-33175, Gradignan Cedex, France}
\shorttitle{Dual States of B0540-69}
\shortauthors{Marshall et~al.}

\begin{abstract}

We report the discovery of a large, sudden, and persistent increase in the
spin-down rate of B0540-69, a young pulsar in the Large Magellanic Cloud,
using observations from the \S\ and \X\ satellites.
The relative increase in the spin-down rate $\dot\nu$ of $36\%$ 
is unprecedented for B0540-69.
No accompanying change in the spin rate is seen, and
no change is seen in the pulsed X-ray emission from B0540-69 following
the change in the spin-down rate.
Such large relative changes in the spin-down rate are seen
in the recently discovered class of ``intermittent pulsars'', and we
compare the properties of B0540-69 to such pulsars.
We consider possible changes in the magnetosphere
of the pulsar that could cause such a large change
in the spin-down rate.

\end{abstract}

\keywords{pulsars: individual (PSR B0540-69)}

\section{Introduction}
\label{introduction}

B0540-69 is a young, rotationally powered pulsar in the Large Magellanic Cloud (LMC)
that is similar to the Crab pulsar in many ways. 
With a rotation frequency $\nu$  of almost 20 Hz, it is one
of the most rapidly rotating young pulsars known.
Its spin-down luminosity ($-4\pi^2I\nu\dot\nu$)
of $\sim1.5 \times 10^{38} I_{45}$ erg $s^{-1}$,
where $I_{45}$ is the moment of inertia of the neutron star (NS)
in units of $10^{45}$ g $cm^{2}$,
is also among the largest for all pulsars.
The spin-down rate of a pulsar is often described in terms
of a braking index $n$, in which 
$\dot\nu = -\kappa\nu^{n}$.
The slowdown of B0540-69  
is relatively stable for a young pulsar, 
and it is one of only 8 young pulsars for which a braking index
has been reliably measured (Lyne \etal\ 2015).
It has a characteristic spin-down age ($-\nu/2\dot{\nu}$) of $\sim1600$ years.

Pulsars are remarkably stable rotators, and the deviations
from clocklike precision can provide information about
the structure and processes at work in the neutron star
and its magnetosphere.
Sudden changes in $\nu$ and $\dot\nu$ (``glitches'') 
are occasionally seen in pulsars,
especially those with ages between $10^{3}$ and $10^{5}$ years
(Yu \etal\ 2013).
Glitches are thought to occur in rotationally powered pulsars
when angular momentum is transferred from a more rapidly rotating
component of the NS to the outer crust (e.g., Anderson \& Itoh 1975;
Franco \etal\ 2000).
A different kind of rapid change is seen in some pulsars 
that affects the spin-down rate.
In their examination of long-term monitoring of 366 pulsars,
Lyne \etal\ (2010) found that many pulsars rapidly switch 
between two different spin-down rates.
The time scales for the transitions are quasi-periodic,
with typical time scales of years.
``Intermittent pulsars'' are extreme examples of such pulsars.
They transition from a radio-on state to radio-off state
with a simultaneous change in the spin-down rate.
The best studied example is PSR B1931+24 (Kramer \etal\ 2006;
Young \etal\ 2013) for which multiple transitions on a timescale
of weeks from a
radio bright state with $\dot\nu = -16 \times 10^{-15}$ Hz s$^{-1}$
and a radio quiet state with $\dot\nu = -10.8 \times 10^{-15}$ Hz s$^{-1}$
have been observed.
Similar behavior on longer time scales has been reported for
PSR J1832+0029 (Lorimer \etal\ 2012) and 
PSR J1841-0500 (Camilo \etal\ 2012).
All three pulsars show rapid transitions between two states
with stable spin-down rates, large differences in the spin-down
rates, large changes in the radio flux, and no simultaneous
change in $\nu$.
Unlike most explanations for glitches, models for these state changes 
have emphasized changes in the pulsar's magnetosphere
(e.g., Li \etal\ 2012).
None of the intermittent pulsars have been detected in
the optical, X-ray, or gamma-ray bands, but somewhat similar
behavior has been seen for the gamma-ray pulsar
PSR J2021+4026 (Allafort \etal\ 2013).
A 4\% increase in $\dot\nu$ was seen with a simultaneous
decrease of 18\% in the flux above 100 MeV with a time scale
for the transition of less than a week.
Since there is no radio counterpart, 
any change in radio flux is not known.

B0540-69 has been extensively studied 
since its discovery with the \E\ (Seward \etal\ 1984). 
A small glitch for B0540-69 was reported by Zhang \etal\ (2001)
with a relative change in $\nu$ of $1.9 \times 10^{-9}$
and $\dot\nu$ of $8.5 \times 10^{-5}$.
The reality of this glitch was disputed by Cusumano \etal\ (2003)
and later supported by Livingstone \etal\ (2005).
Ferdman \etal\ (2015) examined 15.8 years of data from \X\ 
and report a second glitch with
a relative change in $\nu$ of $1.6 \times 10^{-9}$
and $\dot\nu$ of $9.3 \times 10^{-5}$.
Both these changes in $\dot\nu$ are orders of magnitude
smaller than we report in this paper.
No other glitches have been reported in the extensive
monitoring of the pulsar.
Optical pulsations are also seen (Mignani \etal\ 2010
and references therein).
Manchester \etal\ (1993) discovered radio pulsations
with an above average luminosity at 640 MHz of 1200 mJy kpc$^{2}$.

In this paper we present and discuss new timing analysis
of B0540-69 as observed with \X\ and \SP\ 
Details of the observations and
the results of the temporal analysis are reported in Section 2,
and interpretations are given in Section 3. 
Finally, Section 4 is a summary.

\section{Observations and Data Reduction}
\label{obs}
B0540-69 was observed with the Proportional Counter Array (PCA) on board
the {\it Rossi X-ray Timing Explorer} (\X) 
(Bradt \etal\ 1993)
during a 12.5-year campaign to monitor the nearby
PSR~\j0537. The final observation was on Dec. 31, 2011.
We concentrate on observations covering the final 140 days of the campaign.
Results from \X\ observations have been reported
by Zhang \etal\ (2001), Livingstone \etal\ (2005)
and Ferdman \etal\ (2015).
The PCA is composed of five co-aligned xenon detectors 
(Proportional Counter Units (PCUs)) with a total area of $\sim6500$ cm$^2$.
Individual PCUs were routinely turned off and on to reduce the
number of high-voltage breakdowns.
Additional observations were made with the X-Ray Telescope (XRT)
instrument (Burrows \etal\ 2005) on the {\sl Swift Gamma-Ray Burst Explorer}
(Gehrels \etal\ 2004) starting in Feb., 2015.
Table \ref{obs_log} is a log of the observations.

Data reduction for both missions followed very similar procedures.
X-ray events were screened to maximize the signal-to-noise ratio,
and then photon arrival times were corrected to the solar system
barycenter with the
FTOOL\footnote{http://heasarc.gsfc.nasa.gov/docs/software/lheasoft/}
{\scshape faxbary} for \X\ and
{\scshape barycorr} for \S\ 
using the {\sl Hubble Space Telescope} position of $\alpha = 05^{h}40^{m}11.202^{s}$,
$\delta = -69^{\circ}19^{\prime}54.17^{\prime\prime}$ (J2000.0)
(Mignani \etal\ 2010).
The mid-point of each observation was chosen as the epoch, and then
the best period was determined by comparing multiple folded light curves 
using {\scshape efsearch}. 
Since the folded light curve is approximately a sine wave
(Cusumano \etal\ 2003), the phase at the epoch was determined 
by fitting a sine wave to the folded light curve produced
with {\scshape efold}
and using the phase of the peak of the sine wave.
The resulting frequencies and phases were fit to the usual truncated Taylor series
expansion of the phase and its time derivatives
$\nu$, $\dot\nu$, and $\ddot\nu$.

Uncertainties quoted in this paper are given at the 90\% confidence level 
unless otherwise noted.
For multi-parameter fits, the other parameters are allowed to vary
when calculating the limits.

\subsection{{\sl RXTE}/PCA Observations}
\label{pca}

The PCA is sensitive to X-rays in the 2 to 60 keV band with
moderate ($\Delta E /E \sim18\%$) resolution.  Each event is time-tagged
on the spacecraft with an accuracy better than $5 \times 10^{-5}$ s
(Rots \etal\ 1998). To improve the signal-to-noise ratio, only events
in the first xenon layer in the energy range of $3-20$ keV were included.

Figure~\ref{joint_residuals} shows the frequency residuals for the last 14
\X\ observations relative to the best-fit ephemeris model 
for the first 12 of the observations.
The value of $\ddot\nu$ is set to $3.249 \times10^{-21} s^{-3}$,
the best-fit value for observations between Oct. 2010 and Dec. 2011.
The model, whose parameters are given in Table~\ref{ephemeris},
provides a very good fit to the first 12 observations with rms phase residuals
of $\sim$1\%, but the final two observations deviate dramatically.
A linear fit to the frequencies for the final two observations 
indicates that $\dot\nu$ has changed
by $-6.0 \times10^{-11}$ Hz s$^{-1}$ 
with a statistical uncertainty of $0.8 \times10^{-11}$ Hz s$^{-1}$
at some time between the observations
on Dec. 3 and Dec. 17.
This is an increase of $32\pm4\%$ in the rate at which 
the pulsar's spin is slowing.
This value is consistent with, but less accurate than,
the value determined in Section 2.2 using both \S\ and \X\ data. 
We note that our value for $\ddot\nu$ is $\sim14\%$ lower than the values reported
by Ferdman \etal\ (2015) for longer intervals of \X\ data.
Reprocessing the \X\ data 
using the Ferdman \etal\ values for $\ddot\nu$
demonstrates that revising $\ddot\nu$
would have a negligible effect on our results and would
not change our conclusions.

If the sudden change in $\dot\nu$ were due to a glitch, then a simultaneous
change in $\nu$ would be expected.
Surveys of glitches in other pulsars find that
the amplitude of a glitch in $\nu$ correlates with the amplitude
of the glitch in $\dot\nu$ for pulsars in general (Espinoza \etal\ 2011)
and for the Crab Pulsar in particular (Lyne \etal\ 2015).
The correlation for pulsars in general would indicate an accompanying glitch 
in B0540-69 larger than $\sim1 \times10^{-5}$ Hz, and the correlation
for the Crab would indicate an accompanying glitch of $\sim1 \times10^{-3}$ Hz.
There is no indication of an accompanying change in $\nu$, and we
place limits on its possible size by assuming an instantaneous glitch with no
recovery and extrapolating the fit to the post-glitch frequencies
back toward the final pre-glitch observation.
With these assumptions the largest glitch would occur immediately 
after the final pre-glitch observation with a size of $1.2 \times10^{-5}$ Hz.
Later glitches would be smaller until at 200 ks after final pre-glitch observation
there would be no glitch in $\nu$.

\begin{figure}
\includegraphics[angle=270.0,scale=0.5]{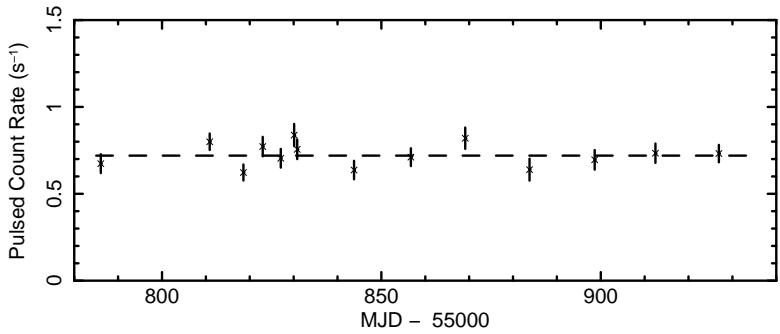}
\caption{
The pulsed count rate with 1-$\sigma$ uncertainties for the \X\ observations.
\label{pulsed_rate}}
\end{figure}

Large glitches in $\dot\nu$ typically decay away on time scales from
about a day to weeks.
Since the new value for $\dot\nu$ is consistent with that later seen with \S\
(Section \ref{xrt}),
there is also no indication of such a recovery.
We conclude that the sudden change in $\dot\nu$ is almost certainly
a state transition rather than a glitch.
We will refer to the state before the transition as the
``low spin-down'' state and the state after the transition as the
``high spin-down'' state.
There is also no indication of a change in the pulsed X-ray
emission from B0540-69. Figure~\ref{pulsed_rate}
shows the average pulsed count rate ($r_{pulse} = A \times C / t_{exp} / N_{PCU}$, 
where A is the relative sine wave amplitude, 
C is the total number of counts detected,
$t_{exp}$ is the exposure time, and $N_{PCU}$ is the average number of PCUs 
that were on during the observation).
The average pulsed count rate for the two observations after the transition
($0.733 \pm 0.086\ s^{-1}$)
is consistent with the average for the first 12 observations
($0.718 \pm 0.035\ s^{-1}$).
Further the folded light curves for these two observations show no significant change.

To confirm the continued nominal performance of the PCA 
including the final two observations, we processed
the same observations for PSR~\j0537\ following the procedures given
in Marshall \etal\ (2004). After the glitch near MJD 55815, the final 12
observations are very well fit using the standard form for the ephemeris
with an rms phase residual of $\sim1\%$.
We conclude that the unexpected values of $\nu$ for B0540-69 are not
due to any instrumental effect.

\subsection{{\sl Swift}~XRT Observations}
\label{xrt}
\begin{figure}
\includegraphics[angle=0.0,scale=0.6]{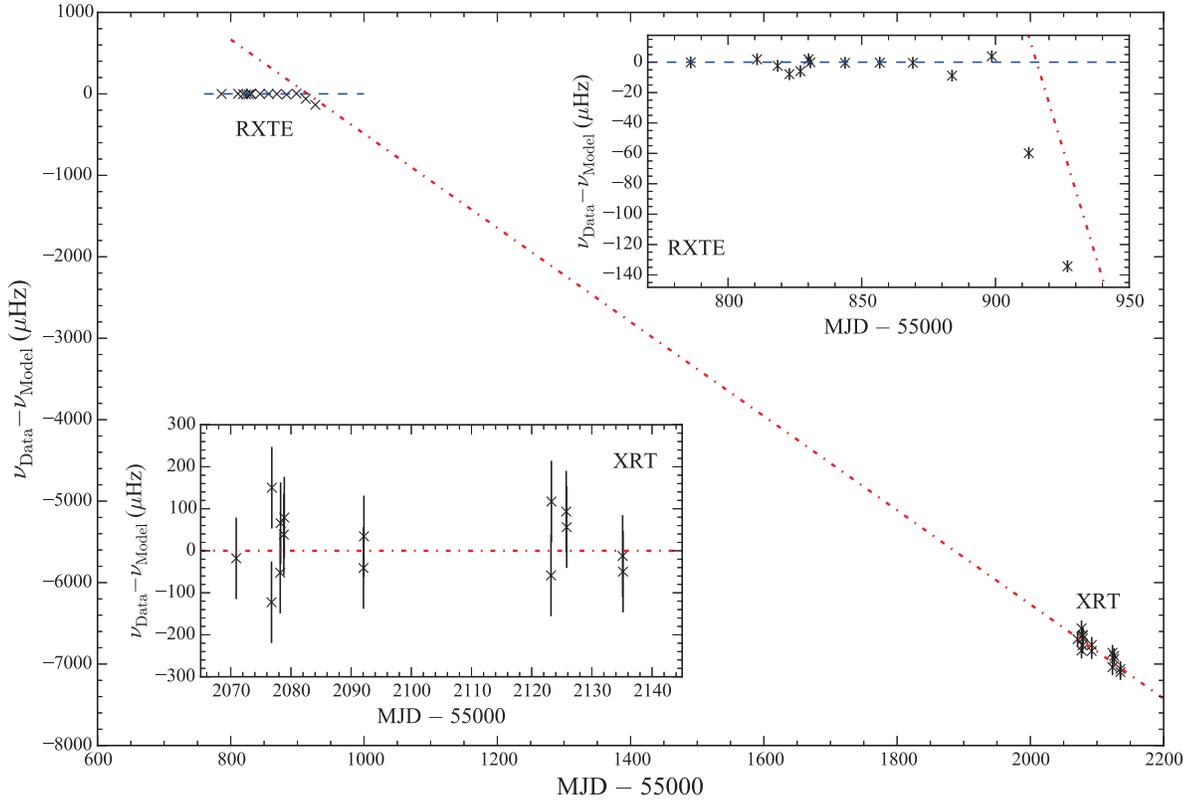}
\caption{The frequency residuals for the \X\ and XRT observations 
relative to the ephemeris model (the dark blue dashed line)
determined with the \X\ observations before the state transition. 
The dash-dotted red line shows the best-fit ephemeris for the XRT data.
One-$\sigma$ uncertainties are shown, but they
are smaller than the symbols for the \X\ data.
\label{joint_residuals}}
\end{figure}

The X-ray Telescope started observations of B0540-69 on Feb. 17, 2015,
as part of a Target of Opportunity campaign. 
The XRT is a focusing X-ray telescope with a CCD detector with an effective
bandpass of 0.3 - 10 keV. 
All observations were made using the Window Timing (WT) mode in which the central
200 CCD columns are continously read out, providing a time resolution
of 1.7 ms.
The XRT data were reduced with the standard software ({\scshape xrtpipeline}
v0.13.1) applying the default filtering and screening criteria 
({\scshape HEASoft 6.16}), using the 20140709 update to the XRT CALDB files
and the 20150428 update to the clock correction file.
The phase and frequency were measured 
for each continuous viewing interval (a ``snapshot'').
Each of the \S\ entries in Table \ref{obs_log} except the first consists
of two snapshots separated by about 96 minutes (the duration of a \S\ orbit).
The first entry has a single snapshot.
We obtained the best ephemeris by determining a model for the closely
spaced observations and then refining the model as more widely separated
observations were added.
All the models assumed a value of $3.249 \times10^{-21} s^{-3}$ for $\ddot\nu$
based on measurements with \X\ before the state transition in 2011.
This parameter makes a very small contribution to the model because the duration
of the current \S\ campaign is only 65 days.
The best-fit model was confirmed using a large grid search of possible
combinations of $\nu$ and $\dot\nu$.
The parameters of the best-fit model and the 90\% confidence
uncertainties are given in Table \ref{ephemeris}.
The model provides a good fit to the data with rms phase
residuals of $\sim$3\%. 
Figure \ref{joint_residuals}, which shows the frequency residuals relative
to the \X\ ephemeris before the transition,
displays the overall history of $\nu$.

We used two methods to determine $\dot\nu$ after the state transition.
The first method uses the best-fit model for the XRT data
(Table \ref{ephemeris}) with a value of 
$-2.52871 \pm 0.00008 \times10^{-10} s^{-2}$ at the XRT epoch.
After adjusting by $-3.33 \times10^{-13} s^{-2}$ to account for the
effect of $\ddot\nu$, this is a change in $\dot\nu$
from the value before the transition (Table \ref{ephemeris})
of $-0.66882 \pm 0.00008 \times10^{-10} s^{-2}$.

The second method compares the average value of $\nu$ in the final two
\X\ observations ($19.72601295$ Hz $\pm 4.6 \times10^{-6}$ Hz at MJD 55919.71)
with the best-fit value for the XRT data of
$19.70077383$ Hz $\pm 4 \times10^{-8}$ Hz.
The resulting $\dot\nu$ is $-2.5230 \pm 0.0005 \times10^{-10}$ s$^{-2}$
at MJD 56498.62.
After adjusting by $1.62 \times10^{-13} s^{-2}$ to account for the
effect of $\ddot\nu$, the result is $-2.5214 \pm 0.0005 \times10^{-10}$ s$^{-2}$
at the XRT epoch, which is larger than result from the first method by
$7.3 \pm 0.5 \times10^{-13}$ s$^{-2}$.
This discrepancy is shown in Figure~\ref{joint_residuals} in which
the extrapolation of the XRT ephemeris back to the final two
\X\ observations predicts a value for $\nu$ that is
$7.3 \times 10^{-5} s^{-1}$ larger than that observed.
The uncertainty of the extrapolated values due to the uncertainty
in the model's $\dot\nu$ is only $8 \times10^{-7}$ Hz.
The discrepancy suggests that the evolution of the high spin-down state
is more complicated than our simple model.
The second method indicates a change in $\dot\nu$
from the value before the transition 
of $-0.6615 \pm 0.0005 \times10^{-10} s^{-2}$.

Since the value of $\ddot\nu$ after the state transition is uncertain
and the statistical uncertainties are small,
we use the value of the adjustments due to $\ddot\nu$ to estimate the uncertainty.
The relative change in $\dot\nu$ is then $35.9\pm0.2\%$
and $35.5\pm0.1\%$ for the first and second methods respectively.
We adopt $35.7\pm0.4\%$,
the average of the two methods with an uncertainty
that encompasses the individual error bars,
as our best estimate of the relative
change in $\dot\nu$.

The folded light curve is similar to that reported from other
X-ray observations (e.g., Cusumano \etal\ 2003).
Although the actual light curve is more complicated,
a model of a constant plus a Gaussian provides a good description of the XRT data.
The best-fit standard deviation for the Gaussian is $0.174 \pm 0.022$,
which is consistent with a fit to the folded light curve for the 
\X\ data before the state transition.

\section{Discussion}
\label{discussion}

B0540-69 has been extensively monitored by numerous
observatories in the more than 30 years since its discovery
including {\sl Ginga} observations spanning 4.4 years
(Deeter \etal\ 1999) and the 12.5 years
of observations with \X.
More than 10 measurements of the braking index for the pulsar,
which require long spans of data, have been reported
(Cusumano \etal\ 2003, who list previous measurements;
Ferdman \etal\ 2015).
Values for the braking index range from 1.81
(Zhang \etal\ 2001) to 2.74 (\"{O}gelman \& Hasinger 1990).

None of these studies found a sudden change in $\dot\nu$ comparable
to the one reported here. 
The glitches in $\dot\nu$ reported by Zhang \etal\ (2001)
and Ferdman \etal\ (2015) are
$\sim$2000 times smaller,
and knowledge of the pulsar's phase is maintained
throughout the 12.5 years of \X\ monitoring until
the final two observations.
When discovered with the \E\ (Seward \etal\ 1984),
B0540-69 had a $\dot\nu$ of $-1.900 \times 10^{-10}$ Hz s$^{-1}$,
which indicates that the pulsar was in the low spin-down state
in 1979 and 1980.
The lack of published values of $\dot\nu$ indicative of
the high spin-down state suggests that B0540-69 remained
in the low spin-down state the vast majority of the time,
if not all the time, until the transition in late 2011.

The sudden change in the spin-down rate is most likely
associated with a change in the magnetosphere of the NS.
One possibility is a global change in the
conductivity of the magnetosphere.
Li \etal\ (2012) developed solutions for pulsar
magnetospheres with finite resistivity that
mitigate some of the limitations of the vacuum
and force-free models.
The dependence of the spin-down luminosity was
calculated for a wide range of the conductivity
parameter $\sigma$ (expressed in terms of the
pulsar angular rotation rate $\Omega$)
and inclination angle $\alpha$ using a three-dimensional
numerical code.
For broad ranges of $\alpha$
(0$^{\circ}$ to 90$^{\circ})$ and the conductivity 
($(\sigma/\Omega)^2$ from 0.04 to 4.0),
a 36\% increase in the luminosity 
(and by extension in $\mid\dot\nu\mid$) can be achieved
with an increase in $\sigma$ of less than a factor of 6.
There are maximum values of $(\sigma/\Omega)^2$, 
ranging from 40 at $\alpha = 0^{\circ}$ to 1.3 at $\alpha = 90^{\circ}$ 
above which it is no longer possible
to explain an increase in the luminosity with
an increase in $\sigma$.
Since for large values of $(\sigma/\Omega)$, 
the component of the electric field parallel to the magnetic field
$E_{\parallel}\ \propto \sigma^{-1}$ 
(Kalapotharakos \etal\ 2014),
an increase in $\sigma$ may reduce the
gamma-ray flux from the pulsar.
The long-term light curve of B0540-69 in the energy range
of 200 MeV to 100 GeV
(Fermi-LAT Collaboration 2015) shows no indication
of change in the gamma-ray flux with an upper limit of 30\%.
Detailed modelling, such as that done by 
Brambilla \etal\ (2015) for other Fermi pulsars,
is needed to understand the implication of this constraint
on parameters for B0540-69, but this is beyond the scope
of this paper.

Another possible explanation for the spin-down transition
is a change in torque due to a change in the plasma outflow
from the polar caps.
Kramer \etal\ (2006) suggested this to explain the state
transitions in the intermittent pulsar PSR B1931+24.
Unlike PSR B1931+24, B0540-69 appears to be radio bright
in its low spin-down state, but its radio flux may
significantly increase or its pulse shape may change 
in the high spin-down state.
Following Young \etal\, we estimate the change
in the charge density $\rho_{plasma}$ as
$7.1 \times 10^{5} \Delta\dot\nu \nu^{-0.5} \dot\nu_{low}^{-0.5}$ C m$^{-3}$,
assuming a pulsar radius of 10 km and 
I$_{45}$ of $1$.
This estimate of 0.78 C m$^{-3}$ is close to the
Goldreich-Julian charge density for B0540-69 of 1.09 C m$^{-3}$.
An increase in $\rho_{plasma}$ predicts an increase
in the radio flux, which can be tested with radio observations
of the high spin-down state.
It also predicts a reduction in the braking index
as the relative importance of magnetic dipole radiation decreases.

A change in the amount of open poloidal magnetic flux
would also change the spin-down luminosity, which is expected
to be proportional to $r_{open}^{-2}$ where $r_{open}$
is the radius beyond which the magnetic field lines become open
(Contopoulos 2007).
An 36\% increase in the spin-down luminosity with no change
in $\nu$ would
require a 15\% decrease in $r_{open}$.

\section{Summary}

The spin-down rate of B0540-69 increased suddenly by 36\%
(a change in $\dot\nu$ of $-6.7 \times10^{-11}$ Hz s$^{-1}$)
between Dec. 3 and Dec. 17, 2011.
Observations with \S\ in 2015 show that this change in spin-down
rate has persisted for more than 3 years. 
Such a large persistent change in the spin-down rate has never
been reported for B0540-69.
There is no indication of a simultaneous change in the pulse shape
in X-rays, the pulsed X-ray luminosity, or the total gamma-ray luminosity.
We interpret this change as a transition between
two stable states of the pulsar, similar to state transitions
in other pulsars. 
B0540-69 extends the class of state changing pulsars to include a very
young and luminous example. 
The transitions in B0540-69 appear to be rare with the low spin-down state
probably lasting more than 30 years.
Future work can test predicted changes in
the radio flux and the braking index in the high spin-down state.

\medskip
We gratefully acknowledge the anonymous referee for suggestions that improved
the paper.
This work made use of data supplied by the UK Swift Science Data Centre at
the University of Leicester and the High Energy Astrophysics Science
Archive Research Center, provided by NASA's Goddard Space Flight Center.

\begin{deluxetable}{llccr}
\tabletypesize{\footnotesize}
\tablecaption{Observing Log\label{obs_log}}
\tablewidth{0pt}
\tablehead{
\colhead{Satellite} &
\colhead{Obs. ID} &
\colhead{Date} &
\colhead{MJD} &
\colhead{$T_{exp}$} \\
\colhead{}  &
\colhead{}  &
\colhead{(1)}  &
\colhead{(1)}  &
\colhead{(2)}  \\
}
\startdata
\X\ & 96023-01-17 & 2011 Aug 13 & $55786$ & $6.6$ \\
\X\ & 96023-01-42 & 2011 Sep 06 & $55810$ & $7.3$ \\
\X\ & 96023-01-43 & 2011 Sep 14 & $55818$ & $7.5$ \\
\X\ & 96023-01-44 & 2011 Sep 18 & $55822$ & $6.9$ \\
\X\ & 96023-01-20 & 2011 Sep 23 & $55827$ & $6.9$ \\
\X\ & 96023-01-45 & 2011 Sep 26 & $55830$ & $7.2$ \\
\X\ & 96023-01-46 & 2011 Sep 26 & $55830$ & $6.3$ \\
\X\ & 96023-01-21 & 2011 Oct 09 & $55843$ & $6.7$ \\
\X\ & 96023-01-22 & 2011 Oct 22 & $55856$ & $6.9$ \\
\X\ & 96023-01-23 & 2011 Nov 04 & $55869$ & $6.9$ \\
\X\ & 96023-01-24 & 2011 Nov 18 & $55883$ & $7.3$ \\
\X\ & 96023-01-25 & 2011 Dec 03 & $55898$ & $6.6$ \\
\X\ & 96023-01-26 & 2011 Dec 17 & $55912$ & $7.1$ \\
\X\ & 96023-01-19 & 2011 Dec 31 & $55926$ & $7.8$ \\
\S\ & 00033603002 & 2015 Feb 17 & $57070$ & $1.2$ \\
\S\ & 00033603004 & 2015 Feb 23 & $57077$ & $2.1$ \\
\S\ & 00033603005 & 2015 Feb 25 & $57078$ & $1.8$ \\
\S\ & 00033603006 & 2015 Feb 25 & $57078$ & $2.4$ \\
\S\ & 00033603007 & 2015 Mar 11 & $57092$ & $2.1$ \\
\S\ & 00033603008 & 2015 Apr 11 & $57123$ & $2.3$ \\
\S\ & 00033603009 & 2015 Apr 13 & $57125$ & $2.3$ \\
\S\ & 00033603010 & 2015 Apr 23 & $57135$ & $2.5$ \\
\enddata
\tablecomments{
(1)~At the start of the observation;
(2)~Exposure time in ks
}
\end{deluxetable}

\begin{deluxetable}{lcc}
\tabletypesize{\footnotesize}
\tablecaption{Ephemeris Parameters\label{ephemeris}}
\tablewidth{0pt}
\tablehead{
\colhead{Parameter} &
\colhead{Pre-transition 2011} &
\colhead{2015} \\
\colhead{}  &
\colhead{(1)} &
\colhead{(1)} \\
}
\startdata
Epoch (Modified Julian Date)\dotfill& 55892.352587464& 57077.537772223 \\
Phase\dotfill& 0.000 (15)& 0.000 (23)\\
$\nu$ (Hz)\dotfill& 19.72655182 (1)& 19.70077383 (4)\\
$\dot\nu$ ($10^{-10} s^{-2}$)\dotfill& -1.86322 (2)& -2.52871 (8)\\
$\ddot\nu$ ($10^{-21} s^{-3}$)\dotfill& 3.249 (fixed)& 3.249 (fixed)\\
\enddata
\tablecomments{
(1)~The numbers in parentheses are the 90\% confidence errors
quoted in the last digit.
}
\end{deluxetable}

\clearpage

\end{document}